\documentclass[pra,aps,twocolumn,showpacs]{revtex4}
\usepackage{}
\usepackage{mathrsfs}
\usepackage{amsfonts}
\usepackage{txfonts}
\usepackage{amssymb}
\usepackage{graphicx}
\usepackage{bm}
\newcommand{\ket}[1]{|#1\rangle}
\newcommand{\bra}[1]{\langle #1|}

\begin{document}
\title{Composite nonadiabatic holonomic quantum computation}
\author{G. F. Xu$^{1,2}$, P. Z. Zhao$^1$, T. H. Xing$^1$, Erik Sj\"{o}qvist$^{2}$ and D. M. Tong$^{1}$\footnote{Email: tdm@sdu.edu.cn}}
\affiliation{$^1$Department of Physics, Shandong University, Jinan
250100, China \\
$^2$Department of Physics and Astronomy, Uppsala University, Box 516, Se-751 20 Uppsala, Sweden}
\date{\today}

\begin{abstract}
Nonadiabatic holonomic quantum computation has robust feature in suppressing control errors because of its holonomic feature. However, this kind of robust feature is challenged since the usual way of realizing nonadiabatic holonomic gates introduces errors due to systematic errors in the control parameters. To resolve this problem, we here propose a composite scheme to realize nonadiabatic holonomic gates. Our scheme can suppress systematic errors while preserving holonomic robustness. It is particularly useful when the evolution period is shorter than the coherence time. We further show that our composite scheme can be protected by decoherence-free subspaces. In this case, the strengthened robust feature of our composite gates and the coherence stabilization virtue of decoherence-free subspaces are combined.
\end{abstract}
\maketitle
\date{\today}

\section{Introduction}
Quantum computation can solve many problems, such as
factoring large integers \cite{Shor} and searching unsorted databases \cite{Grover}, much faster than classical computation. To achieve the advantages of quantum computation, realizing a universal set of quantum gates with sufficiently high fidelities is needed.
However, high-fidelity gates are always difficult to be realized.
One main practical obstacle in realizing high-fidelity gates is control errors. Especially, if the evolution period is shorter than the coherence time, control errors become a dominant obstacle. Although control errors can be reduced by improving the calibration of apparatus, such suppression method is not always practical. Thus, finding a method to suppress control errors is particularly useful for any quantum computer implementation.

Nonadiabatic holonomic quantum computation is a promising method to suppress control errors \cite{Sjoqvist,Xu}. Nonadiabatic holonomic gates depend only on evolution paths of a quantum system but not on evolution details, and thus they are robust against certain control errors.
Although the proposal of nonadiabatic holonomic quantum computation was only recently proposed, it has immediately received increasing attentions due to both the robust feature and the high-speed implementation. So far, many schemes of realizing nonadiabatic holonomic gates have been put forward for various physical systems \cite{Abdumalikov,Feng,Mousolou,Mousolou1,
Zhang1,Arroyo,Zu,Xu1,Xu2,Liang,Zhang2,Zhou,Xue,Sjoqvist1,Pyshkin,Song,Xue1,Wang,Herterich,Sun,Xu3,Sjoqvist2}. Particularly, nonadiabatic holonomic quantum computation has been
experimentally demonstrated with circuit QED, NMR, and nitrogen-vacancy (NV) center in diamond \cite{Abdumalikov,Feng,Arroyo,Zu}.

The most important feature of nonadiabatic holonomic quantum computation is its robustness in suppressing control errors. To realize nonadiabatic holonomic gates, systems with at least three dimensional Hilbert space are needed. As is well known, controlling methods for higher-dimensional systems are more complicated than that for qubits because more controlling parameters need to be manipulated. Accordingly, in the presence of imperfect calibrations, more imperfect controlling parameters, i.e., systematic errors, are encountered. Generally, more systematic errors induce more infidelities unless the gates are specially designed. To resolve the above problem, we here propose a composite scheme to realize nonadiabatic holonomic gates. Our scheme not only preserves the holonomic robustness but also suppresses systematic errors. It is particularly useful when the evolution period is shorter than the coherence time. We further show that our composite scheme can be protected by decoherence-free subspaces. As a result, the strengthened robust feature of our composite gate and the coherence stabilization virtue of decoherence-free subspaces are combined.

\section{The composite scheme}
Before proceeding further, we briefly explain how nonadiabatic holonomy arises in unitary evolution.
Consider an $N-$dimensional quantum system exposed to the Hamiltonian $H(t)$.
Suppose there is an $L-$dimensional subspace $\mathcal{S} (0)$ forming the computational state space and spanned by the orthonormal basis vectors
$\{ \ket{\phi_k(0)} \}_{k=1}^L$, and the state of the system is initially in this subspace. The evolution operator is a nonadiabatic holonomy acting on the subspace $\mathcal{S} (0)$ if the following two conditions are satisfied:
$\textrm{(i)} \ \sum_{k=1}^L\ket{\phi_k (\tau)} \bra{\phi_k (\tau)} =
\sum_{k=1}^L \ket{\phi_k (0)} \bra{\phi_k (0)}$,
and
$\textrm{(ii)} \ \bra{\phi_k (t)}H(t)\ket{\phi_l (t)}=0, \ k,l = 1, \ldots ,L$, where $\tau$ is the evolution period and $\ket{\phi_k(t)} = {\bf T} \exp{[-i\int_0^tH(t')dt']} \ket{\phi_k(0)}$, with ${\bf T}$ being time ordering. The first condition guarantees that the action on the computational subspace is unitary; the second condition ensures that the dynamical phase vanishes and the evolution becomes purely geometric.

\subsection{One-qubit gates}
\begin{figure}[htbp]
\begin{center}
\includegraphics[width=8.5cm, height=4.2cm]{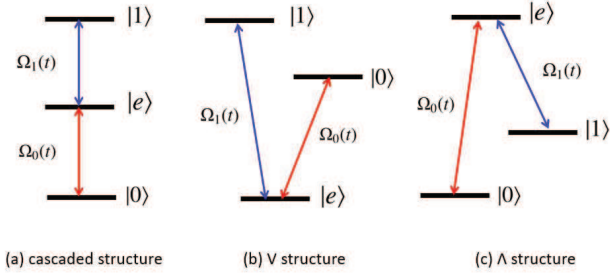}
\end{center}
\caption{(color online). Three possible level configurations of the three-level system with the two driving laser fields. (a) Cascaded structure, (b) V structure, and (c) $\Lambda$ structure. $\ket{0}$ and $\ket{1}$ are computational states, and $\ket{e}$ is an ancillary state.
$\Omega_j(t)$ is the Rabi frequency of laser field driving the transition $\ket{j}\leftrightarrow\ket{e}$.}\label{fig1}
\end{figure}

We now elucidate the physical model of the one-qubit gates. Consider a three-level system with eigenstates $\ket{0}$,   $\ket{1}$, and $\ket{e}$, where $\ket{0}$ and $\ket{1}$ are computational states, and $\ket{e}$ is an ancillary state. The transitions $\ket{0}\leftrightarrow\ket{e}$ and $\ket{1}\leftrightarrow\ket{e}$ are respectively coupled by two resonant laser fields, as shown in Fig. \ref{fig1}.
In this case, the Hamiltonian can be written as
\begin{eqnarray}
H(t)=H_0+{\bm \mu}\cdot[{\bf E}_0(t)+{\bf E}_1(t)],
\end{eqnarray}
where $H_0=-\omega_{e0}\ket{0}\bra{0}-\omega_{e1}\ket{1}\bra{1}$ with $\omega_{ej}=\omega_e-\omega_j$ being the bare Hamiltonian, ${\bm \mu}$ is the electric dipole operator, and ${\bf E}_j(t)={\bm e}_jg_j(t)\cos\nu_jt$ with ${\bm e}_j$ being the polarization, $g_j(t)$ being the envelope function, and $\nu_j$ being the oscillation frequency.
By using the rotating frame and the rotating wave approximation, the effective Hamiltonian reads
\begin{eqnarray}
H(t)=\Omega_0(t)e^{i\phi_0}\ket{0}\bra{e}+\Omega_1(t)e^{i\phi_1}\ket{1}\bra{e}+h.c., \label{h1}
\end{eqnarray}
where $\Omega_0(t)$, $\Omega_1(t)$ are Rabi frequencies, $\phi_0$, $\phi_1$ are laser phases, and $h.c.$ represents Hermitian conjugation. We assume that the Rabi frequencies $\Omega_0(t)=\Omega(t)\cos\frac{\theta}{2}$ and $\Omega_1(t)=\Omega(t)\sin\frac{\theta}{2}$, where $\Omega(t)$ is the real-valued envelope and $\theta$ is a time-independent parameter representing the relative strengths of the two Rabi frequencies.
Then the Hamiltonian $H(t)$ can be rewritten as
\begin{eqnarray}
H(t)=\Omega(t)\big(e^{i\phi_0}\ket{b_{\theta,\phi}}\bra{e}
+e^{-i\phi_0}\ket{e}\bra{b_{\theta,\phi}}\big), \label{h2}
\end{eqnarray}
where
$\ket{b_{\theta,\phi}}=\cos\frac{\theta}{2}\ket{0}+\sin\frac{\theta}{2}e^{i\phi}\ket{1}$ is the bright state,
with phase $\phi=\phi_1-\phi_0$.
Given this form of Hamiltonian, the dark state
$\ket{d_{\theta,\phi}}=\sin\frac{\theta}{2}\ket{0}-\cos\frac{\theta}{2}e^{i\phi}\ket{1}$
decouples from the dynamics all the time.

The Hamiltonian $H(t)$ can be generated by a variety of quantum systems, such as trapped ions, circuit QED, nitrogen-vacancy center, and so on. In the following, we use the Hamiltonian $H(t)$ to realize arbitrary one-qubit nonadiabatic holonomic gates with the ability to suppress systematic errors.

To realize such robust gates, we use the method of composite pulses to strengthen the robust features of nonadiabatic holonomic gates. Specifically, we first use the Hamiltonian $H(t)$ to realize the elementary gate
\begin{eqnarray}
U_{\theta,\phi}=-i\ket{e}\bra{e}+i\ket{b_{\theta,\phi}}\bra{b_{\theta,\phi}}
+\ket{d_{\theta,\phi}}\bra{d_{\theta,\phi}}, \label{elementarygate}
\end{eqnarray}
and then use the elementary gate $U_{\theta,\phi}$ to realize composite gates not only preserving holonomic robustness but also suppressing systematic errors.

To better show the realization of the desired composite gates, we need to know the form of systematic errors in our case. For the Hamiltonian $H(t)$, the most unstable controlling parameters are Rabi frequencies. Thus the dominant source of systematic errors is pulse strength, i.e., Rabi frequency, errors \cite{Low,Ivanov}. This kind of errors occurs when the strength of the driving field used to induce transition deviates from its nominal value by an unknown fraction. In the presence of pulse strength errors, the Hamiltonian $H(t)$ turns into
\begin{eqnarray}
H^\prime(t)=(1+\epsilon_0)\Omega_0(t)e^{i\phi_0}\ket{0}\bra{e}
+(1+\epsilon_1)\Omega_1(t)e^{i\phi_1}\ket{1}\bra{e}
+h.c.,
\label{herror1}
\end{eqnarray}
where $\epsilon_0$ and $\epsilon_1$ represent the unknown time-independent fractions of the two laser fields.
Since the Rabi frequencies $\Omega_0(t)$ and $\Omega_1(t)$ have the same envelope $\Omega(t)$, the Hamiltonian $H^\prime(t)$ can be rewritten as
\begin{eqnarray}
H^\prime(t)=(1+\epsilon)\Omega(t)\big(e^{i\phi_0}\ket{b_{\theta^\prime,\phi}}\bra{e}
+e^{-i\phi_0}\ket{e}\bra{b_{\theta^\prime,\phi}}\big),\label{herror2}
\end{eqnarray}
where
\begin{eqnarray}
\epsilon=\sqrt{(1+\epsilon_0)^2\cos^2\frac{\theta}{2}
+(1+\epsilon_1)^2\sin^2\frac{\theta}{2}}-1,\label{epsilon}
\end{eqnarray}
and the error-affected bright state $\ket{b_{\theta^\prime,\phi}}$ is
\begin{eqnarray}
\ket{b_{\theta^\prime,\phi}}=\cos\frac{\theta^\prime}{2}\ket{0}
+\sin\frac{\theta^\prime}{2}e^{i\phi}\ket{1},\label{berror}
\end{eqnarray}
with
\begin{eqnarray}
\theta^\prime=2\arctan\big(\frac{1+\epsilon_1}{1+\epsilon_0}\tan\frac{\theta}{2}\big). \label{thetaerror}
\end{eqnarray}
Correspondingly, the error-affected dark state is
\begin{eqnarray}
\ket{d_{\theta^\prime,\phi}}=\sin\frac{\theta^\prime}{2}\ket{0}
-\cos\frac{\theta^\prime}{2}e^{i\phi}\ket{1}.
\end{eqnarray}
Equation (\ref{herror2}) clearly shows that the pulse strength errors of the two laser fields affect not only the absolute value of the envelope $\Omega(t)$, but also the angle $\theta$. Specifically, $\Omega(t)$ and $\theta$ turns into $(1+\epsilon)\Omega(t)$ and $\theta^\prime$, respectively. In the following, we assume that $\mid\epsilon_0\mid, \mid\epsilon_1\mid\ll1$, which  implies that $\mid\epsilon\mid\ll1$ and $\theta^\prime\approx\theta$.

After knowing the form of systematic errors, we now show the realization of the desired composite gates. We start by considering the following composite gate
\begin{eqnarray}
U_{\theta,\phi}U_{\theta,\phi}=-\ket{e}\bra{e}-\ket{b_{\theta,\phi}}\bra{b_{\theta,\phi}}
+\ket{d_{\theta,\phi}}\bra{d_{\theta,\phi}}, \label{cu1}
\end{eqnarray}
where $U_{\theta,\phi}$ is the elementary gate described by Eq. (\ref{elementarygate}).
The elementary gate $U_{\theta,\phi}$ is generated by the Hamiltonian $H(t)$ according to
\begin{eqnarray}
U_{\theta,\phi}=e^{-i\int_{T_1}^T{\Omega^{\prime\prime}(t)(\ket{b_{\theta,\phi}}\bra{e}+h.c.)}dt}
e^{-i\int_0^{T_1}{\Omega^\prime(t)(i\ket{b_{\theta,\phi}}\bra{e}+h.c.)}dt},
\end{eqnarray}
where $T_1$ is an intermediate time, $T$ is the total time, and $\int_0^{T_1}{\Omega^\prime(t)}dt=\int_{T_1}^T{\Omega^{\prime\prime}(t)}dt=\frac{\pi}{2}$.
From the above equation, one can see that the whole evolution is divided into two segments. For the first segment $t\in[0, T_1]$, the phase $\phi_0$ of $H(t)$ is chosen to be $\frac{\pi}{2}$, while for the second segment $t\in[T_1, T]$, the phase $\phi_0=0$. The whole evolution of the gate $U_{\theta,\phi}$ forms a loop in the Grassmannian $\mathcal {G}(3;2)$, i.e., the space of two-dimensional subspaces of a three-dimensional Hilbert space, and each segment completes half of this loop.
It can be verify that $U_{\theta,\phi}$ is a one-qubit nonadiabatic holonomic gate because the two conditions  $\textrm{(i)}$ and $\textrm{(ii)}$ mentioned in the first paragraph of this section are satisfied \cite{Herterich}.
Accordingly, the composite gate $U_{\theta,\phi}U_{\theta,\phi}$ is also a one-qubit nonadiabatic holonomic gate, and thus has holonomic robustness.

To investigate the ability of $U_{\theta,\phi}U_{\theta,\phi}$ to suppress pulse strength errors, we consider
the error-affected composite gate
\begin{eqnarray}
U^\prime_{\theta,\phi}U^\prime_{\theta,\phi}
=U_\theta{U_{\Omega(t)}}
={U_{\Omega(t)}}U_\theta, \label{cuerror1}
\end{eqnarray}
where $U_\theta$ and $U_{\Omega(t)}$ are defined as
\begin{eqnarray}
U_\theta=-\ket{e}\bra{e}-\ket{b_{\theta^\prime,\phi}}\bra{b_{\theta^\prime,\phi}}
+\ket{d_{\theta^\prime,\phi}}\bra{d_{\theta^\prime,\phi}}, \label{utheta}
\end{eqnarray}
and
\begin{eqnarray}
{U_{\Omega(t)}}&=&
e^{-i\frac{\epsilon\pi}{2}(\ket{b_{\theta^\prime,\phi}}\bra{e}+h.c.)}
e^{i\frac{\epsilon\pi}{2}(i\ket{b_{\theta^\prime,\phi}}\bra{e}+h.c.)}
\nonumber\\
&&e^{i\frac{\epsilon\pi}{2}(\ket{b_{\theta^\prime,\phi}}\bra{e}+h.c.)}
e^{-i\frac{\epsilon\pi}{2}(i\ket{b_{\theta^\prime,\phi}}\bra{e}+h.c.)}.
\end{eqnarray}
By iteratively using the Baker-Campbell-Hausdorff (BCH) relation
$e^Be^A=\exp{(B+A+\frac{1}{2}[B,A]+\cdots)}$
to the operator ${U_{\Omega(t)}}$, we find that the first-order terms of $\epsilon$, i.e., main deviation of $\Omega(t)$, are cancelled out and  ${U_{\Omega(t)}}$ is reduced to
\begin{eqnarray}
{U_{\Omega(t)}}=I+O(\epsilon^2), \label{uomegaerror}
\end{eqnarray}
where $I$ represents the identity operator of the three-level system.
Thus, the error-affected gate $U^\prime_{\theta,\phi}U^\prime_{\theta,\phi}$ can be written as
\begin{eqnarray}
U^\prime_{\theta,\phi}U^\prime_{\theta,\phi}=-\ket{e}\bra{e}
-\ket{b_{\theta^\prime,\phi}}\bra{b_{\theta^\prime,\phi}}
+\ket{d_{\theta^\prime,\phi}}\bra{d_{\theta^\prime,\phi}}+O(\epsilon^2).
\label{uprimeuprime}
\end{eqnarray}
The above equation clearly shows that the detrimental effect associated with  $\Omega(t)\rightarrow(1+\epsilon)\Omega(t)$ is suppressed by implementing $U_{\theta,\phi}U_{\theta,\phi}$. However, the sensitivity to the error-induced change $\theta\rightarrow\theta^\prime$ is not changed.

To further suppress the detrimental effect associated with $\theta\rightarrow\theta^\prime$, we next consider the composite gate
\begin{eqnarray}
U_{\theta,\phi}U_{\theta,\phi}U_{\pi-\theta,\phi}U_{\pi-\theta,\phi}
=\ket{e}\bra{e}+\exp{\big[i(\pi-2\theta)\sigma_{\phi+\frac{\pi}{2}}\big]}. \label{cu2}
\end{eqnarray}
Here, $U_{\pi-\theta,\phi}$ turns into $U_{\theta,\phi}$ by replacing $\pi-\theta$ with $\theta$, and the notation $\sigma_\alpha$ is
$\sigma_\alpha=\cos\alpha\sigma_x+\sin\alpha\sigma_y$,
where $\alpha$ is a phase, and $\sigma_x$ and $\sigma_y$ respectively represent Pauli $X$ and $Y$ operators acting on the logical subspace. It can be verified that the composite gate
$U_{\theta,\phi}U_{\theta,\phi}U_{\pi-\theta,\phi}U_{\pi-\theta,\phi}$ has nonadiabatic holonomic robustness.

We can get error-affected $U_{\theta,\phi}U_{\theta,\phi}U_{\pi-\theta,\phi}U_{\pi-\theta,\phi}$ by using Eq. (\ref{uprimeuprime}) and it reads
\begin{eqnarray}
U^\prime_{\theta,\phi}U^\prime_{\theta,\phi}
U^\prime_{\pi-\theta,\phi}U^\prime_{\pi-\theta,\phi}
=U_\theta{U}_{\pi-\theta}+O(\epsilon^2), \label{uerror4}
\end{eqnarray}
where $U_{\pi-\theta}$ turns into $U_\theta$ defined in Eq. (\ref{utheta}) by replacing $\pi-\theta$ with  $\theta$. Equation (\ref{uerror4}) shows that the composite gate $U_{\theta,\phi}U_{\theta,\phi}U_{\pi-\theta,\phi}U_{\pi-\theta,\phi}$ preserves the ability to suppress the detrimental effect associated with $\Omega(t)\rightarrow(1+\epsilon)\Omega(t)$. To show that this gate can also suppress the detrimental effect associated with $\theta\rightarrow\theta^\prime$, we rewrite $U_\theta{U}_{\pi-\theta}$ as
\begin{eqnarray}
U_\theta{U}_{\pi-\theta}=
\ket{e}\bra{e}+U^L_{\theta}U^L_{\pi-\theta}. \label{uthetautheta}
\end{eqnarray}
where $U^L_{\theta}$ is defined as
\begin{eqnarray}
U^L_{\theta}=-\ket{b_{\theta^\prime,\phi}}\bra{b_{\theta^\prime,\phi}}
+\ket{d_{\theta^\prime,\phi}}\bra{d_{\theta^\prime,\phi}},
\end{eqnarray}
which turns into $U^L_{\pi-\theta}$ by replacing $\theta$ with $\pi-\theta$. Since both operators $U^L_{\theta}$ and $U^L_{\pi-\theta}$ are logical operators acting on the computational subspace $\{\ket{0},\ket{1}\}$, we can have
\begin{eqnarray}
U^L_{\theta}&=&-\sin\theta^\prime\sigma_\phi
-\cos\theta^\prime\sigma_z, \nonumber\\
U^L_{\pi-\theta}&=&-\sin(\pi-\theta)^\prime\sigma_\phi
-\cos(\pi-\theta)^\prime\sigma_z, \label{uu}
\end{eqnarray}
where $(\pi-\theta)^\prime$ equals to $\pi-\theta$ in the absence of errors and $\sigma_z$ is Pauli $Z$ operator acting on the computational subspace. By using Eq. (\ref{uu}) and the relation
$\sigma_{\phi}=i\sigma_z\sigma_{\phi+\frac{\pi}{2}}$,
the logical operator $U^L_{\theta}U^L_{\pi-\theta}$ can be rewritten as
\begin{eqnarray}
U^L_{\theta}U^L_{\pi-\theta}=\exp\big\{i[(\pi-\theta)^\prime-\theta^\prime]\sigma_{\phi+\frac{\pi}{2}}\big\}.
\label{ulul}
\end{eqnarray}
To proceed further, we need to investigate the value of $(\pi-\theta)^\prime-\theta^\prime$.
Without loss of generality, the error-affected angle $(\pi-\theta)^\prime$ can be rewritten as
\begin{eqnarray}
(\pi-\theta)^\prime=\pi-\theta^{\prime\prime},
\end{eqnarray}
where $\theta^{\prime\prime}$ is defined as an angle which makes the above equation valid.
According to Eq. (\ref{thetaerror}), the value of $\theta^{\prime\prime}$ reads
\begin{eqnarray}
\theta^{\prime\prime}=2\arctan\big(\frac{1}{1+x}\tan\frac{\theta}{2}\big),
\end{eqnarray}
where
\begin{eqnarray}
x=\frac{\epsilon_1-\epsilon_0}{1+\epsilon_0}.
\end{eqnarray}
Small errors in the Rabi frequencies implies $\mid{x}\mid\ll1$ and $\theta^{\prime\prime}$ can be expanded as
\begin{eqnarray}
\theta^{\prime\prime}=\theta-\frac{2\tan\frac{\theta}{2}}{1+\tan^2\frac{\theta}{2}}{x}+\cdots.
\label{thetatheta}
\end{eqnarray}
Similarly, $\theta^\prime$ can be expanded as
\begin{eqnarray}
\theta^\prime=\theta+\frac{2\tan\frac{\theta}{2}}{1+\tan^2\frac{\theta}{2}}{x}+\cdots. \label{thetatheta1}
\end{eqnarray}
By using Eqs. (\ref{thetatheta}) and (\ref{thetatheta1}), we can readily get $(\pi-\theta)^\prime-\theta^\prime$. Then, substituting $(\pi-\theta)^\prime-\theta^\prime$ into Eq. (\ref{ulul}), we can have
\begin{eqnarray}
U^L_{\theta}U^L_{\pi-\theta}
=\exp\big[i(\pi-2\theta)\sigma_{\phi+\frac{\pi}{2}}\big]+O(x^2). \label{ulul1}
\end{eqnarray}
Since the first-order terms of $x$ are canceled out, the detrimental effect associated with  $\theta\rightarrow\theta^\prime$ is suppressed. According to Eqs. (\ref{uerror4}), (\ref{uthetautheta}) and (\ref{ulul1}), we can finally get
\begin{eqnarray}
&&U^\prime_{\theta,\phi}U^\prime_{\theta,\phi}
U^\prime_{\pi-\theta,\phi}U^\prime_{\pi-\theta,\phi}= \nonumber\\
&&\ket{e}\bra{e}+\exp{\big[i(\pi-2\theta)\sigma_{\phi+\frac{\pi}{2}}\big]}+O(\epsilon^2)+O(x^2). \label{ufinal}
\end{eqnarray}
Equations (\ref{cu2}) and (\ref{ufinal}) clearly show that the gate $U_{\theta,\phi}U_{\theta,\phi}U_{\pi-\theta,\phi}U_{\pi-\theta,\phi}$
can simultaneously suppress the detrimental effects associated with both $\Omega(t)\rightarrow(1+\epsilon)\Omega(t)$ and $\theta\rightarrow\theta^\prime$, and thus is robust against pulse strength errors. Since $U_{\theta,\phi}U_{\theta,\phi}U_{\pi-\theta,\phi}U_{\pi-\theta,\phi}$ also has nonadiabatic holonomic feature, it suppresses systematic errors while preserving holonomic robustness.

To show the suppression ability more clear, we calculate the fidelity of the gate by using the formula
\begin{eqnarray}
F_{U,V}=\frac{\mid{Tr}(U^\dag{V})\mid}{{Tr}(U^\dag{U})},
\end{eqnarray}
where $U$ is the desired gate and $V$ is the error-affected gate. Through calculation, the fidelity can be simply written as
\begin{eqnarray}
F=1-O(\bar{\epsilon}^4).
\end{eqnarray}
where $O(\bar{\epsilon}^4)$ represents that the fidelity is fourth-order error dependence.
It noteworthy that the fidelity of the usual nonadiabatic holonomic gate is second-order error dependence. Thus, the advantage of our composite gate in suppressing pulse strength errors is obvious.

\subsection{Two-qubit gate}

In the basis $\{\ket{0}, \ket{1}\}$, the gate  $U_{\theta,\phi}U_{\theta,\phi}U_{\pi-\theta,\phi}U_{\pi-\theta,\phi}$ is equivalent to the logical gate $\exp{[i(\pi-2\theta)\sigma_{\phi+\frac{\pi}{2}}]}$, which is sufficient to realize arbitrary one-qubit gates. However, to realize universal quantum computation, a nontrivial two-qubit gate is also needed. We now demonstrate how to realize a two-qubit composite nonadiabatic holonomic gate with the ability to suppress  systematic errors.

To start with, we elucidate the physical model of the two-qubit gate. To be compatible with the realized one-qubit gates, we consider two three-level systems. Previous researches have constructed Hamiltonians with three-level structures by appropriately manipulating two three-level systems. See, for example, the Hamiltonians in Refs. \cite{Sjoqvist,Zu}. Although both physical systems and controlling methods are different in these researches, the  Hamiltonians can be uniformly written as
\begin{eqnarray}
\mathcal {H}(t)=\Omega_{jk}e^{i\phi_{jk}}\ket{jk}\bra{a}+\Omega_{lm}e^{i\phi_{lm}}\ket{lm}\bra{a}+h.c.,
\label{htwoqubit}
\end{eqnarray}
where $\ket{jk}$ and $\ket{lm}$ are two orthogonal computational states, $\ket{a}$ is an ancillary state which is orthogonal to the computational subspace, $\Omega_{jk}$, $\Omega_{lm}$ are Rabi frequencies, and $\phi_{jk}$, $\phi_{lm}$ are phases, with $j,k,l,m\in\{0, 1\}$. According to Eqs. (\ref{h1}) and (\ref{htwoqubit}), we can see that $\mathcal {H}(t)$ has the same structure as the one-qubit Hamiltonian $H(t)$.
Moreover, the models of systematic errors for both $\mathcal {H}(t)$ and $H(t)$ also have the same form. Set $\Omega_{lm}$ to zero and then in the presence of pulse strength errors, the Hamiltonian $\mathcal {H}(t)$ turns into
\begin{eqnarray}
\mathcal {H}_{\epsilon_{jk}}(t)=(1+\epsilon_{jk})\Omega_{jk}e^{i\phi_{jk}}\ket{jk}\bra{a}+h.c.,
\end{eqnarray}
where $\epsilon_{jk}$ is the error fraction of the Rabi frequency $\Omega_{jk}$.
The above equation shows that the pulse strength errors only affects the Rabi frequency $\Omega_{jk}$, which  turns into $(1+\epsilon_{jk})\Omega_{jk}$. Compared with $H^\prime(t)$ in Eq. (\ref{herror1}), which has two deviations $\Omega(t)\rightarrow(1+\epsilon)\Omega(t)$ and $\theta\rightarrow\theta^\prime$, Hamiltonian $\mathcal {H}_{\epsilon_{jk}}$ only has one deviation $\Omega_{jk}\rightarrow(1+\epsilon_{jk})\Omega_{jk}$. Thus, we avoid to tackle the problem associated with the change in the angle $\theta$. Accordingly, we can only use the way, in which we suppress the detrimental effect associated with $\Omega(t)\rightarrow(1+\epsilon)\Omega(t)$ in the one-qubit case, to realize the desired two-qubit gate. So, we consider the following two-qubit composite gate
\begin{eqnarray}
U_{jk}U_{jk}=-\ket{a}\bra{a}-\ket{jk}\bra{jk}+\sum_{hn\neq{jk}}\ket{hn}\bra{hn},
\end{eqnarray}
where $U_{jk}$ represents the elementary gate which can be realized by the Hamiltonian $\mathcal {H}(t)$,
\begin{eqnarray}
U_{jk}=e^{-i\int_{\tau_{jk}}^{T_{jk}}{\Omega^{\prime\prime}_{jk}(\ket{jk}\bra{a}+\ket{a}\bra{jk})}dt}
e^{-i\int_0^{\tau_{jk}}{\Omega^\prime_{jk}(i\ket{jk}\bra{a}-i\ket{a}\bra{jk})}dt},
\end{eqnarray}
with $h,n,j,k\in\{0,1\}$ and $\int_0^{\tau_{jk}}\Omega^\prime_{jk}dt=\int_{\tau_{jk}}^{T_{jk}}{\Omega^{\prime\prime}_{jk}}dt=\frac{\pi}{2}$. By calculation, the elementary gate $U_{jk}$ can be written as
\begin{eqnarray}
U_{jk}=-i\ket{a}\bra{a}+i\ket{jk}\bra{jk}+\sum_{hn\neq{jk}}\ket{hn}\bra{hn}.
\end{eqnarray}
By checking the holonomic conditions $\textrm{(i)}$ and $\textrm{(ii)}$, the composite gate $U_{jk}U_{jk}$ is a nonadiabatic holonomic gate and thus has holonomic robustness.
Similar to case of the one-qubit gate $U_{\theta,\phi}U_{\theta,\phi}$, we can verify that
the composite gate $U_{jk}U_{jk}$ has the ability to suppress systematic errors,
\begin{eqnarray}
U^\prime_{jk}U^\prime_{jk}=-\ket{a}\bra{a}-\ket{jk}\bra{jk}+\sum_{hn\neq{jk}}\ket{hn}\bra{hn}+O(\epsilon^2_{jk}),
\end{eqnarray}
where $U^\prime_{jk}$ is the error-affected $U_{jk}$. It can be verified that the fidelity of the gate $U_{jk}U_{jk}$ is also fourth-order error dependence,
\begin{eqnarray}
F_{jk}=1-O(\epsilon_{jk}^4).
\end{eqnarray}
In addition, the composite gate $U_{jk}U_{jk}$ is a nontrivial two-qubit gate. So, we have realized a nontrivial two-qubit gate which suppresses systematic errors while preserving holonomic robustness.

\section{Suppression of decoherence}
We have proposed a scheme which not only suppresses systematic errors but also preserves holonomic robustness. This means our scheme has strengthened robustness against control errors. We now demonstrate that our composite scheme can be protected by decoherence-free subspaces. Accordingly, the strengthened robust feature in suppressing control errors and the coherence stabilization virtue of decoherence-free subspaces are combined.

The main reason our composite scheme can be protected from decoherence is that our composite scheme does not depend on specific physical systems. As long as the Hamiltonians have three-level structures, our composite scheme can be realized.
It has been shown that realizing Hamiltonians with three-level structures in decoherence-free subspaces is feasible \cite{Liang,Xue1}. Thus, our scheme can be protected by decoherence-free subspaces.

To be more specific, we focus a system consisting of an array of two-level ions \cite{Liang}.
Consider three two-level ions interacting collectively with a dephasing environment. For such interaction, there exists a three-dimensional decoherence-free subspace
\begin{eqnarray}
\mathcal{S}_1 = \text{Span} \big\{ \ket{100}, \ket{001}, \ket{010} \big\}.
\end{eqnarray}
Denote the computational basis elements as $\ket{0}_L = \ket{100}$, $\ket{1}_L = \ket{001}$, and the remaining vector as ancillae $\ket{a}_L=\ket{010}$.
To realize the one-logical-qubit gates, two lasers acting on ion $1$ ($2$) are tuned to frequencies $\omega_0+(\nu+\delta)$ with phase $\varphi_1$ ($\varphi_2$), and two lasers acting on ion $2$ ($3$) are tuned to frequencies $\omega_0+(\nu-\delta)$ with phase $\varphi_2^\prime$ ($\varphi_3$). The effective Hamiltonian of the three ions reads
\begin{eqnarray}
H_{1}=
\frac{\eta^2}{\delta}\big(\mid\Omega_{12}\mid^2e^{i\varphi_{12}}\ket{a}_L\bra{0}_L
-\mid\Omega_{23}\mid^2e^{i\varphi_{23}}\ket{a}_L\bra{1}_L+h.c.\big), \nonumber\\
\end{eqnarray}
where $\Omega_{12}$, $\Omega_{23}$ are Rabi frequencies, $\varphi_{12}=\varphi_1-\varphi_2$, and $\varphi_{23}=\varphi_2^\prime-\varphi_3$. Comparing the Hamiltonian $H_1$ with the Hamiltonians $H(t)$ and $\mathcal {H}(t)$ in Eqs. (\ref{h1}) and (\ref{htwoqubit}), we find that these Hamiltonians have the same structure, i.e., three-level structure. Thus, we can use Hamiltonian $H_1$ to realize arbitrary one-logical-qubit composite gates with both nonadiabatic holonomic robustness and ability to suppress systematic errors in $\mathcal{S}_1$.

In the case of two-logical-qubit gate, six two-level ions interacting collectively with a dephasing environment are considered. Ions $1,2,3$ ($4,5,6$) represent the logical qubit $1$ ($2$). The decoherence-free subspace is written as
\begin{eqnarray}
\mathcal{S}_2 = \text{Span} \big\{ \ket{00}_L, \ket{01}_L, \ket{10}_L, \ket{11}_L, \ket{a_1}_L, \ket{a_2}_L \big\},
\end{eqnarray}
where $\ket{a_1}_L=\ket{101000}$ and $\ket{a_2}_L=\ket{000101}$.  Two lasers acting on ion $3$ ($4$) are tuned to frequencies $\omega_0+(\nu+\delta)$ with phase $\varphi_3^\prime$ ($\varphi_4$), and two lasers acting on ion $3$ ($6$) are tuned to frequencies $\omega_0+(\nu-\delta)$ with phase $\varphi_3^{\prime\prime}$ ($\varphi_6$). The resulting effective Hamiltonian reads
\begin{eqnarray}
H_{2}&=&
\frac{\eta^2}{\delta}\big[\mid\Omega_{34}\mid^2e^{i\varphi_{34}}(\ket{a_1}_L\bra{00}_L+\ket{a_2}_L\bra{11}_L)
\nonumber\\
&&-\mid\Omega_{36}\mid^2e^{i\varphi_{36}}(\ket{a_1}_L\bra{01}_L+\ket{a_2}_L\bra{10}_L)+h.c.\big],
\end{eqnarray}
where $\Omega_{34}$, $\Omega_{36}$ are Rabi frequencies, $\varphi_{34}=\varphi_3^\prime-\varphi_4$, and $\varphi_{36}=\varphi_3^{\prime\prime}-\varphi_6$. We can see that the subspaces $\text{Span} \{ \ket{00}_L, \ket{01}_L, \ket{a_1}_L, \}$ and $\text{Span} \{ \ket{10}_L, \ket{11}_L, \ket{a_2}_L \}$
are independent and the Hamiltonian $H_2$ has three-level structure in each subspace. Thus, although the Hamiltonian $H_2$ is complicated, we can still use Hamiltonian $H_2$ to realize two-logical-qubit composite gate with both nonadiabatic holonomic robustness and ability to suppress systematic errors in $\mathcal{S}_2$.

\section{Conclusions}
The most important feature of nonadiabatic holonomic quantum computation is its robustness in suppressing control errors. However, this robust feature is challenged because more systematic errors are encountered in realizing nonadiabatic holonomic gates. To resolve this problem, we have proposed a scheme which suppresses systematic errors while preserving holonomic robustness. We sequentially implement four elementary gates to realize the one-qubit composite gates and two elementary gates to realize the two-qubit composite gate. Our scheme does not depend on specific physical systems, and it works as long as the Hamiltonians have three-level structures. Our scheme is particularly useful when the evolution period is shorter than the coherence time. We also show that our composite scheme can be protected by decoherence-free subspaces and, as a result, the strengthened robust feature in suppressing control errors and the coherence stabilization virtue of decoherence-free subspaces are combined.

It is interesting to note that using compositions of several gates to suppress systematic errors is a topic that has attracted considerable attention in the case of dynamical gates \cite{Levitt}. However, whether one can use such method to suppress the systematic errors of holonomic gates has not been addressed until now. Here, we not only answer the above question but also give a specific realization of the desired composite gates. As we know, besides pulse strength errors, there also exist other kinds of systematic errors, e.g. detuning errors, which can have detrimental effects on the holonomic gates. So, in the future, finding a way to suppress other kinds of systematic errors in holonomic schemes is interesting and worth paying attention to. Our scheme is promising in experimental implementation, which may shed light on the applications of nonadiabatic holonomic quantum computation.

\section*{Acknowledgments}
G.F.X. acknowledges support from the National Natural Science Foundation of China through Grants No. 11547245 and No. 11605104, and from the Future Project for Young Scholars of Shandong University through Grant No. 2016WLJH21. P.Z.Z. acknowledges support from the National Natural Science Foundation of China through Grant No. 11575101. E.S. acknowledges financial support from the Swedish Research Council (VR) through Grant No. D0413201. D.M.T. acknowledges support from the National Basic Research Program of China through Grant No. 2015CB921004.


\begin{thebibliography}{99}
\bibitem{Shor}P. W. Shor, SIAM J. Comput. {\bf 26}, 1484 (1997).
\bibitem{Grover}L. K. Grover, Phys. Rev. Lett. {\bf 79}, 325 (1997).
\bibitem{Sjoqvist}E. Sj\"{o}qvist, D. M. Tong, B. Hessmo, M. Johansson, and K.
Singh, New J. Phys. {\bf 14}, 103035 (2012).
\bibitem{Xu}G. F. Xu, J. Zhang, D. M. Tong, E. Sj\"{o}qvist, and L. C. Kwek,
Phys. Rev. Lett. {\bf 109}, 170501 (2012).
\bibitem{Abdumalikov}A. A. Abdumalikov, J. M. Fink, K. Juliusson, M. Pechal,
S. Berger, A. Wallraff, and S. Filipp, Nature (London) {\bf 496},
482 (2013).
\bibitem{Feng}G. R. Feng, G. F. Xu, and G. L. Long, Phys. Rev. Lett. {\bf 110},
190501 (2013).
\bibitem{Mousolou}V. A. Mousolou, C. M. Canali, and E. Sj\"{o}qvist,
New J. Phys. {\bf 16}, 013029 (2014).
\bibitem{Mousolou1}V. A. Mousolou and E. Sj\"{o}qvist,
Phys. Rev. A {\bf 89}, 022117 (2014).
\bibitem{Liang}Z. T. Liang, Y. X. Du, W. Huang, Z. Y. Xue, and H. Yan, Phys. Rev. A {\bf 89}, 062312 (2014).
\bibitem{Zhang1}J. Zhang, L. C. Kwek, E. Sj\"{o}qvist, D. M. Tong, and P.
Zanardi, Phys. Rev. A {\bf 89}, 042302 (2014).
\bibitem{Xu1}G. F. Xu and G. L. Long, Sci. Rep. {\bf 4}, 6814 (2014).
\bibitem{Xu2}G. F. Xu and G. L. Long, Phys. Rev. A {\bf 90}, 022323 (2014).
\bibitem{Arroyo}S. Arroyo-Camejo, A. Lazariev, S. W. Hell, and G. Balasubramanian, Nat. Commun. {\bf 5}, 4870 (2014).
\bibitem{Zu}C. Zu, W. B. Wang, L. He, W. G. Zhang, C. Y. Dai, F. Wang, and L. M. Duan, Nature (London) {\bf 514}, 72 (2014).
\bibitem{Zhang2}J. Zhang, T. H. Kyaw, D. M. Tong, E. Sj\"{o}qvist, and L. C. Kwek,
Sci. Rep. {\bf 5}, 18414 (2015).
\bibitem{Sjoqvist1}E. Sj\"{o}qvist, Int. J. Quantum Chem. {\bf 115}, 1311 (2015).
\bibitem{Xue}Z. Y. Xue, J. Zhou, and Z. D. Wang, Phys. Rev. A {\bf 92}, 022320 (2015).
\bibitem{Zhou}J. Zhou, W. C. Yu, Y. M. Gao, and Z. Y. Xue, Opt. Express {\bf 23}, 14027 (2015).
\bibitem{Pyshkin}P. V. Pyshkin, D. W. Luo, J. Jing, J. Q. You, and L. A. Wu, Sci. Rep. {\bf 6}, 37781 (2016).
\bibitem{Xu3}G. F. Xu, C. L. Liu, P. Z. Zhao, and D. M. Tong, Phys. Rev. A {\bf 92}, 052302 (2015).
\bibitem{Sjoqvist2}E. Sj\"{o}qvist, Phys. Lett. A {\bf 380}, 65 (2016).
\bibitem{Wang}Y. Wang, J. Zhang, C. Wu, J. Q. You, and G. Romero, Phys. Rev. A 94, 012328 (2016).
\bibitem{Song}X. K. Song, H. Zhang, Q. Ai, J. Qiu, and F. G. Deng, New J. Phys. {\bf 18}, 023001 (2016).
\bibitem{Sun}C. F. Sun, G. C. Wang, C. F. Wu, H. D. Liu, X. L. Feng, J. L.
Chen, and K. Xue, Sci. Rep. 6, 20292 (2016).
\bibitem{Xue1}Z. Y. Xue, J. Zhou, Y. M. Chu, and Y. Hu, Phys. Rev. A {\bf 94}, 022331 (2016).
\bibitem{Herterich}E. Herterich and E. Sj\"{o}qvist, Phys. Rev. A {\bf 94}, 052310 (2016).
\bibitem{Low}G. H. Low, T. J. Yoder, and I. L. Chuang, Phys. Rev. A {\bf 89}, 022341 (2014).
\bibitem{Ivanov}S. S. Ivanov and N. V. Vitanov, Phys. Rev. A {\bf 92}, 022333 (2015).
\bibitem{Levitt}M. H. Levitt, Prog. Nucl. Magn. Res. Spectr. {\bf 18}, 61 (1986).
\end{thebibliography}
\end{document}